\documentclass[conference]{IEEEtran}
\IEEEoverridecommandlockouts

\usepackage{amssymb,amsmath,amsfonts,bm,epsfig,graphicx,theorem,latexsym}
\usepackage{rotating,setspace,latexsym,epsf,color,subfigure}
\usepackage[spaces,hyphens]{url}
\usepackage{cite,authblk,bbm,stfloats}
\usepackage{multirow,fixltx2e}
\usepackage[utf8]{inputenc}
\def \n2{{N_0 \over 2}}

\def \h5{\hspace{0.5in}}

\begin{document}
	\IEEEoverridecommandlockouts
	\pagestyle{empty}
	
	\title{Overage and Staleness Metrics for Status Update Systems}
	\author
	
	\author{Peng Zou$^{1}$, Jin Zhang$^{1}$, Xianglin Wei$^{2}$ and Suresh Subramaniam$^{1}$ \\
		\normalsize $^{1}$ECE Department, 
		George Washington University, 
		Washington DC, 20052, USA\\
		\normalsize $^{2}$The 63rd Research Institute, National University of Defense Technology, Nanjing 210007, China \\
		\normalsize {\it pzou94, zhangjin, xianglinwei, suresh@gwu.edu}}
	
	\maketitle
	\begin{abstract}
		Status update systems consist of sensors that take measurements of a physical parameter and transmit them to a remote receiver. Age of Information (AoI) has been studied extensively as a metric for the freshness of information in such systems with and without an enforced hard or soft deadline. In this paper, we propose three metrics for status update systems to measure the ability of different queuing systems to meet a threshold requirement for the AoI. The {\em overage probability} is defined as the probability that the age of the most recent update packet held by the receiver is larger than the threshold. The {\em stale update probability} is the probability that an update is stale, i.e., its age has exceeded the deadline, when it is delivered to the receiver. Finally, the {\em average overage} is defined as the time average of the overage (i.e., age beyond the threshold), and is a measure of the average ``staleness'' of the update packets held by the receiver. We investigate these metrics in three typical status update queuing systems -- M/G/1/1, M/G/1/$2^*$, and M/M/1. Numerical results show the performances for these metrics under different parameter settings and different service distributions. The differences between the average overage and average AoI are also shown. Our results demonstrate that a lower bound exists for the stale update probability when the buffer size is limited. Further, we observe that the overage probability decreases and the stale update probability increases as the update arrival rate increases.
		%{\color{red} Need a statement or two on the observations.}
	\end{abstract}
	
	\section{Introduction}
\pagenumbering{arabic}

 The need for \textit{real-time} communication of status update packets involves maintaining information freshness in many mission-critical applications. Age of Information (AoI) is a metric measuring the freshness of  information at the receiver of a status update system. Since its introduction in \cite{ kaul2012real} for queuing models motivated from vehicular status update systems, the AoI metric has been found useful in numerous applications that require timely availability of information at the receiving end of a communication or remote-controlled system, such as tele-surgery or skill transfer in tactile Internet.

While AoI provides a measure of the freshness of information, it does not account for soft or hard deadlines that applications impose on the age. Motivated by the need to consider deadlines, practitioners have recently considered a threshold for the age and investigated when the threshold is exceeded.
%We note that the use of threshold has been a topic of research in earlier works in the literature on AoI. 
 In \cite{zhang2021aoi}, the authors propose a freshness-aware refreshing scheme in which the cached content items will be refreshed to the up-to-date version upon user request if the AoI exceeds a certain threshold.  In \cite{li2020aoi}, the authors consider scheduling problems at the network edge when each source node has an AoI requirement which is called Maximum AoI Threshold (MAT). A threshold-based ALOHA scheme, in which each terminal attempts transmission with constant probability in each slot when the age exceeds the threshold is studied in \cite{yavascan2021analysis}. In \cite{jiang2020analyzing}, the authors propose a simple threshold policy that achieves the optimum  asymptotically for AoI in a wireless multi-access network with the thresholds explicitly derived. We considered a variant of threshold-based age in our previous paper~\cite{zou2020age} wherein an update packet has a value associated with it and the value decays with time and becomes zero beyond a threshold. %we propose three metrics for status update systems %\textcolor{blue}{to measure the ability of different queuing systems to meet a threshold requirement for the AoI.} \textcolor{red}{I'm not sure what you mean by the previous sentence.}

%As peak AoI is proposed in \cite{costa2016age}, we are interested in studying the AoI over the threshold. {\color{red} Was this the first paper that studied peak age?}

In this paper, we expand upon the initial work on threshold-based age and present three metrics of relevance in various status update systems. Given a threshold for the instantaneous age, the metrics are defined as follows. The {\em overage probability} is the probability that the age of the most recent packet currently held by the receiver is larger than the threshold. The {\em stale update probability} is the probability that a delivered update is stale, i.e., its age has exceeded the threshold, when it is delivered to the receiver. This metric is relevant when the receiving system takes action immediately upon the receipt of the packet. Finally, the {\em average overage} is defined as the time average of the overage (i.e., age beyond the threshold), and is a measure of the average ``staleness'' of the update packet held by the receiver. We investigate these three metrics in three typical status update queuing systems, namely, M/G/1/1, M/G/1/$2^*$, and M/M/1, and obtain closed-form expressions for these metrics. We present numerical results for these metrics under different parameter settings and service time distributions. We also show a comparison between the average overage and the average AoI.   

The overage probability has been studied in a few scenarios previously. 
%Previous work in \cite{pan2020information,pan2021coding,hsu2020age,seo2019outage,franco2019analysis,kim2020sensing,hu2021status} have components related to our view on overage probability. 
In \cite{pan2021coding}, the authors use bounded AoI to denote the probability that the AoI is under a given threshold, which is the opposite of our overage probability metric. The authors of \cite{Inoue2019} obtain the distribution of AoI under different queuing schemes, which can in turn be used to derive the overage probability.  In \cite{Devassy2018,Zheng2018,hsu2020age,seo2019outage,franco2019analysis,kim2020sensing,hu2021status,hou2021,zhang2021,Champati2019}, the probability that the age at the receiver is larger than the given threshold is denoted as violation probability or outage probability. While much research has been done for the threshold model in AoI optimization problems, little attention has been paid to the definition and derivation of threshold-based metrics. In this paper, we call this probability the overage probability, and analyze it for the three queuing systems; we note that the overage probability has not been analyzed for the M/G/1/1 and M/G/1/2$^{*}$ systems before. Further, to the best of our knowledge, the other two metrics we propose here have not been considered before. Thus, our contributions in this paper include: (a) the proposal of two new metrics, (b) the derivation of expressions for all metrics for three commonly studied queuing systems, and (c) numerical results exploring various trade-offs. 

%{\color{red} Write a para outlining the contributions of this paper and giving the paper organization. Can you try to make the paper fit in 6 pages eventually?}
%In our work, we have these following contribution comparing to the previous work.
%\begin{itemize}
%\item We make definition for two new metrics based on the overage probability which indicates the time average overage and the probability of stale update.
%\item We did the analysis for these three metrics in general queue scheme including M/GI/1/1 and M/GI/1$2^*$. For the M/M/1 queue which has been studied in \cite{hu2021status}, we show the analysis for the average overage and stale update probability as well.
%\item Our numerical results compare the performance of these three metrics for different queue schemes and compare the average overage with average general AoI.
%\end{itemize}

	\section{System Model}
\begin{figure}[!t]
	\centering{
		\includegraphics[totalheight=0.05\textheight]{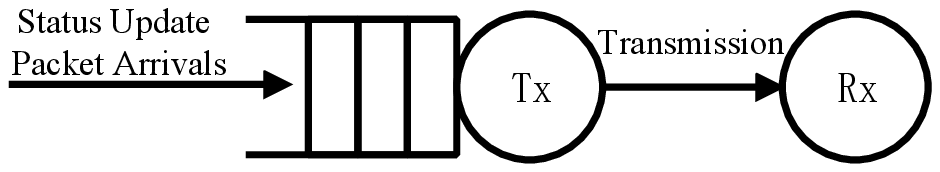}}
	\caption{\sl Status update packets arriving to a single server transmission queue.}
	\label{fig:1} 
	\vspace{-0.25in}
\end{figure}

We consider a point-to-point communication system with a transmitter sending status updates from a single source to a receiver as shown in Fig. \ref{fig:1}. The update packets arrive at the transmitter following a Poisson process with an arrival rate $\lambda$. A packet may be discarded in the queuing phase based on buffer size and the packet dropping policy; those that are not discarded enter the server and are received by the receiver after a random service time. Here, we only index those packets that are received by the receiver and use $S_i$ to denote the service time for the $i$th packet. $S_i$ is assumed to be iid with pdf $f_S(s)$. In this paper, we consider M/GI/1/1, M/GI/1/$2^*$, and M/M/1 queuing schemes. In M/GI/1/1, there is no buffer and packets arriving in busy state are discarded. In M/GI/1/$2^*$, there is a single data buffer and an arriving packet replaces the packet in the buffer (if any).
In M/M/1, there is infinite buffer space for the packets with first come first serve discipline. The first two systems with no and limited buffer have been extensively studied recently \cite{kaul2012status,costa2016age,CKam2016MILCOM, Yates2019, champati2020minimum} because it is known that limited buffering helps decrease AoI. We also study the M/M/1 system here because it was originally studied in \cite{kaul2012real} and our analysis enables a comparison of these new metrics with the AoI.
%The average AoI measures the long-term average age of update packets at a receiver and does not shed light into the usefulness of these updates when updates have to meet certain deadlines. We introduce the value of information with deadline in our previous work which take the usefulness of these status update packets into consideration for different systems \cite{zou2020age}. 

Inspired by \cite{li2020aoi} and \cite{pan2021coding}, we introduce a deadline/threshold $H$ for the age of the update packets. Unlike the hard threshold assumption that instantaneous age is not allowed to exceed the threshold in \cite{li2020aoi}, our threshold is assumed as a soft threshold as in \cite{pan2021coding}. Based on a given threshold, we introduce three metrics -- {\em overage probability}, {\em average overage}, and {\em stale update probability} -- and formally define them below.

Our metrics are closely related to the peak age. As discussed in \cite{costa2016age} for peak age, the AoI reaches a local maximum value every time before a packet is delivered. Based on the queue states and system time of the packets, peak age is different for different received packets. Therefore, if a fixed threshold for instantaneous AoI is given, some received packets' peak age will exceed the threshold while some peak ages will stay under the threshold. At the same time, the AoI is supposed to decrease at the instant when the packet is delivered to the receiver. At this instant, the instantaneous AoI is equal to the system time of this packet. 
\subsection{Overage Probability}
The {\em overage probability} is the probability that the age of the packet currently held by the receiver is larger than the threshold. Let $\varepsilon_i$ denote the time period during which the instantaneous age is larger than the threshold $H$ during the interval between the delivery of packet $i-1$ and packet $i$ (also called the inter-departure period between packet $i-1$ and packet $i$). Let $Y_i$ denote the inter-departure period between packet $i-1$ and packet $i$ and $T_i$ denote the system time (i.e., waiting time plus service time) of packet $i$. Then, the peak age caused by packet $i$ is $\Delta_i^{peak}=T_{i-1}+Y_{i}$, and we have $\varepsilon_i$ as:
\begin{align}\label{epsi}
	\varepsilon_i=\left\{\begin{matrix}
		&0 &\Delta_i^{peak}<H \\ 
		&\Delta_i^{peak}-H  &\Delta_i^{peak}>H,\ T_{i-1}<H \\
		&Y_i& T_{i-1}>H.
	\end{matrix}\right.
\end{align}
Then the overage probability is:
\begin{align}\label{oageP}
	P_{o}=\lim_{T \to \infty} \frac{1}{T}\sum_{i=1}^{n(T)}\varepsilon_i,
\end{align}
where $T$ is the observation period and $n(T)$ is the number of packets delivered to the receiver in observation period $T$.
%{\color{red} Redraw the figure 2 to also illustrate the case when $\epsilon_i = \Delta_i^{peak}-H$ for another packet. You can use this to also show the difference between the two different cases of $Q_i$ when you explain that eqn (6).}
\subsection{Average Overage}
The {\em average overage} is defined as the time average of the update age beyond the threshold, and is a measure of the average ``staleness'' of the update packet held by the receiver. The instantaneous AoI is the difference between the current time and the generation time of the packet held by the receiver: 
\begin{align*} 
	\Delta(t)=t-u(t), 
\end{align*} 
where $u(t)$ is the generation time of the latest packet at the receiver at time $t$. The instantaneous overage is then given by:
\begin{align}\label{oage}
	\Delta^o(t)=\max(\Delta(t)-H,0),
\end{align}
 and the average overage is:
 \begin{align}\label{aoage}
 	\mathbb{E}[\Delta^o]=\lim_{T \to \infty} \frac{1}{T}\int_{0}^{T} \Delta^o(t)dt.
 \end{align}
 \subsection{Stale Update Probability}
Finally we introduce the {\em stale update probability}. This is the probability that an update is stale, i.e., its age has exceeded the deadline, when it is delivered to the receiver. It is clear that packets with system time larger than the threshold are stale since those packets cannot make the age decrease below the threshold. Furthermore, packets that are discarded in the queuing system are also considered to be stale, they are never delivered and can be thought of as having infinite system time. We define $p_d$ as the probability that a packet is delivered to the receiver, i.e., not dropped. Then the stale update probability, $P_s$ is given by:
\begin{align}\label{Ps}
	P_s=1-p_d+{\rm Pr^*}\{T_i>H\}
\end{align}
where ${\rm Pr^*}\{T_i>H\}$ is the probability that a packet is a delivered packet with system time large than threshold.

\begin{figure}[!t]
	\centering{
		%\hspace{-0.3cm} 
		\includegraphics[totalheight=0.21\textheight]{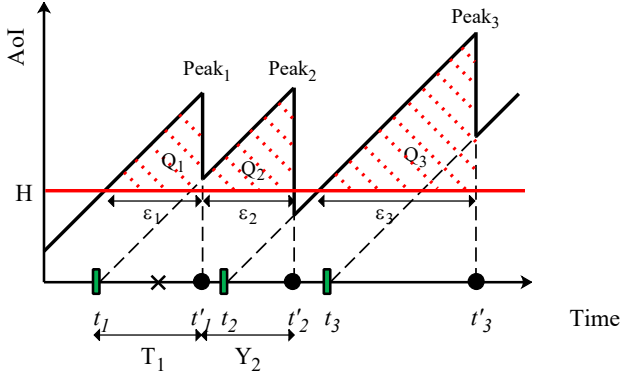}}
	\caption{\sl Illustration of overage probability, stale update probability, and average overage. }
	\label{fig:Overage} 
\end{figure}

The three metrics are illustrated in Fig. \ref{fig:Overage} for an M/GI/1/1 system. At time $t_1$, packet 1 arrives to the system and starts its service. During the service time of packet 1, a packet arrives to the system and is dropped since there is no buffer to store this packet. This time is denoted by a cross in the figure. The service of packet 1 is finished at $t_1'$; packet 2 arrives to the system at $t_2$ and finishes its service at $t_2'$. We can see that packet 1 is a stale update packet whereas packet 2 is not since the system time of packet 2 is smaller than the threshold.

%{\color{red} I suggest that you lightly shade the $Q_i$ areas, as it's not obvious because there are multiple lines. Also, remove the empty space around the section headings sections.}
%{\color{blue} I have changed the figure. If everything is ok, I will remove the question in red and remove the empty space in paper.}
The average overage is the time average area of the age beyond the threshold, and we use $Q_i$ to denote those areas during the inter-departure period between packet $i-1$ and packet $i$. These areas are shown shaded in Fig. \ref{fig:Overage}. From the figure, we can see that $\varepsilon_2$ and $Q_2$ are obtained using the expression for the second case (i.e., $\Delta_i^{peak}>H,\ T_{i-1}<H$) and $\varepsilon_3$ and $Q_3$ are obtained using the expression for the third case (i.e., $T_{i-1}>H$). %{\color{red} Check the previous sentence and ensure it's accurate.} {\color{blue} The sentence looks fine for me.} 
$Q_i$ and $\varepsilon_i$ are related as follows:
\begin{align}\label{Qi}
	Q_i=\left\{\begin{matrix}
		&0 &\Delta_i^{peak}<H \\ 
		&\frac{1}{2}\varepsilon_i^2  &\Delta_i^{peak}>H,\ T_{i-1}<H \\
		&\frac{1}{2}\varepsilon_i^2+\varepsilon_i(T_{i-1}-H)& T_{i-1}>H,
	\end{matrix}\right.
\end{align}
and we can rewrite equation (\ref{aoage}) as:
\begin{align}\label{aoage2}
	\mathbb{E}[\Delta^o]=\lim_{T \to \infty} \frac{1}{T}\sum_{i=1}^{n(T)}Q_i.
\end{align}
If we denote the number of generated packets by the source in observation period $T$ by $N(T)$, we will have $\lambda=\lim_{T \to \infty}\frac{N(T)}{T}$.
%{\color{red} You should have an expectation for $N(T)$ here.}
Therefore, we have:
\begin{align}\label{Po}
	P_o=\lambda\lim_{T \to \infty}\frac{\sum_{i=1}^{n(T)}\varepsilon_i}{N(T)}=\lambda\mathbb{E}[\varepsilon], 
\end{align}\vspace{-0.1in}
\begin{align}\label{Do}
	\mathbb{E}[\Delta^o]=\lambda\lim_{T \to \infty}\frac{\sum_{i=1}^{n(T)}Q_i}{N(T)}=\lambda\mathbb{E}[Q],
\end{align}
where $\mathbb{E}[\varepsilon]$ and $\mathbb{E}[Q]$ are expectations taken over {\em all generated packets} by assuming a value of zero for $\varepsilon$ and $Q$ for dropped packets.
%{\color{red} I don't understand this sentence.}

In the next three sections, we derive expressions for the three metrics for the three queuing systems.
	%{\color{red} Reduce the spacing before and after section heading.}
\section{M/GI/1/1 Queue}\label{partA}
In the $M/GI/1/1$ queuing system, there is a single server and no buffer. Packets that arrive when the server is idle are taken to service immediately and those arriving in busy period are dropped. In view of the renewal structure, we have the following stationary probabilities for each state:
\begin{equation}\label{MG11:pb}
	p_{I}=\frac{1}{\lambda T_{cycle}},\ p_{B}=\frac{\mathbb{E}[S]}{T_{cycle}},
\end{equation}
where $T_{cycle}=\frac{1}{\lambda}+\mathbb{E}[S]$ is the expected length of one renewal cycle; $I$ and $B$ indicate the idle and busy states. Clearly, we have $p_d=p_{I}$ which indicates that only the packets arrive in idle state can be delivered to the receiver. Since there is no waiting period for packets (as there is no buffer), we have $T_i=S_i$, $Y_i=X_i+S_i$ where $X_i$ denotes the idle period before packet $i$ arrives to the system and is an exponentially distributed random variable with coefficient $\lambda$. We now obtain expressions for the metrics. Closed-form expressions may be obtained for specific service time distributions. We show closed-form expressions for the M/M/1/1 system in Appendix.\ref{Amm11}. %Due to page limitation and uncertain of distribution for service time, we show the integration from which the closed form expression can be derived with in the rest of the paper.  
\subsection{Overage probability}
First, we evaluate $\mathbb{E}[\varepsilon]$ which is equal to $\mathbb{E}[\varepsilon|Id]p_{I}$, since the packets arriving in busy state will be dropped in this system. For $\mathbb{E}[\varepsilon|Id]$, based on equation (\ref{epsi}), we have:
\begin{align}\label{EeMG11}
\nonumber	\mathbb{E}[\varepsilon|Id]=\int_{0}^{H}\int_{0}^{\infty}\int_{H-s_i-s_{i-1}}^{\infty}(x+s_{i}+s_{i-1}-H)\\
\nonumber f_{X}(x)f_{S}(s_i)f_{S}(s_{i-1})dxds_ids_{i-1}\\
+\int_{H}^{\infty}\int_{0}^{\infty}\int_{0}^{\infty}(x+s_{i})
f_{X}(x)f_{S}(s_i)f_{S}(s_{i-1})dxds_ids_{i-1}.
\end{align}
Finally, we have overage probability as  $P_o=\lambda\mathbb{E}[\varepsilon|Id]p_{I}$ from equation (\ref{Po}). The more general case of a GI/GI/1/1 system was considered in \cite{Champati2019}.
\subsection{Average Overage}
Similarly, we evaluate $\mathbb{E}[Q|Id]$ based on (\ref{Qi}):
\begin{align}\label{EQMG11}
	\nonumber	\mathbb{E}[Q|Id]=\int_{0}^{H}\int_{0}^{\infty}\int_{H-s_i-s_{i-1}}^{\infty}\frac{1}{2}(x+s_{i}+s_{i-1}-H)^2\\
	\nonumber f_{X}(x)f_{S}(s_i)f_{S}(s_{i-1})dxds_ids_{i-1}\\
\nonumber	+\int_{H}^{\infty}\int_{0}^{\infty}\int_{0}^{\infty}[\frac{1}{2}(x+s_{i})^2+s_{i-1}(x+s_{i})]\\
	f_{X}(x)f_{S}(s_i)f_{S}(s_{i-1})dxds_ids_{i-1}.
\end{align}
Then we have $\mathbb{E}[\Delta^o]=\lambda\mathbb{E}[Q|Id]p_{I}$ from equation (\ref{Do}). 
\subsection{Stale Update Probability}
Since $p_d=p_{I}$, based on equation (\ref{Ps}), we evaluate ${\rm Pr^*}\{T_i>H\}$ as:
\begin{align}
\nonumber	{\rm Pr^*}\{T_i>H\}&=p_{I}{\rm Pr^*}\{T_i>H|Id\}\\&=p_{I}\int_{H}^{\infty}f_{S}(s_i)ds_i=p_{I}(1-F_S(H)),
\end{align}
where $F_S(s)$ is the cumulative distribution function (cdf) of random variable $S_i$.

Therefore, we have $P_s=1-p_{I}F_S(H)$ from equation (\ref{Ps}). 
	\section{M/GI/1/$2^*$ Queue}
In this system, there is a single packet buffer. The server is in either idle or busy state. Packets that arrive in the idle period are served immediately and those that arrive in the busy period are stored in the buffer and take the place of the old packet in the buffer. In view of the renewal structure, we have the following stationary probabilities for each state of the server:
\begin{align}\label{prob:MG12}
	p_{I}=\frac{1}{\lambda T_{cycle}},\ p_{B}=\frac{\mathbb{E}[S]}{T_{cycle}MGF_{S}(\lambda)},
\end{align}
where we use $MGF_{S}(\lambda)$ to denote the moment generating function of the service distribution evaluated at $-\lambda$:
\begin{align}
	MGF_{S}(\lambda)= \mathbb{E}[e^{-\lambda S}],
\end{align} 
where $T_{cycle}=\frac{1}{\lambda}+\frac{\mathbb{E}[S]}{MGF_{S}(\lambda)}$ is the expected length of one renewal cycle.

Since only one packet that arrives during the busy period is served and the others are discarded in the buffer, we define the states $B_{1}$ and $B_2$ as the busy states of the server with zero and one packet waiting in the queue, respectively. The renewal cycle is as follows: after idle period, an arrival happens, and the system enters $B_{1}$ state. Now a time duration of service $S$ starts and if during the service period another arrival occurs, the system turns to $B_{2}$ state. This back-and-forth between $B_1$ and $B_2$ states continues until no packet arrives in one service time. Based on the analysis in our previous work \cite{zou2020age}, we have:
	\begin{align*}
		p_{B_{2}}=p_{B}\left(1+\frac{MGF_{S}(\lambda)-1}{\lambda \mathbb{E}[S]}\right),
	\end{align*}
	and the probability of $B_{1}$ state is $p_{B_{1}}=p_{B}-p_{B_{2}}$.
Clearly we have $p_d=p_{I}+p_{B_{1}}$ for M/GI/1/$2^*$ system. We are now ready to derive the three metrics for this queuing system.  Closed-form expressions the M/M/1/$2^*$ system are shown in Appendix.\ref{Amm12}.
\subsection{Overage probability}
Note that a packet can be delivered to the receiver even if it arrives to the system during busy period. Therefore, we evaluate $\mathbb{E}[\varepsilon|Id]$ and $\mathbb{E}[\varepsilon|B]$ separately to get the expression for $\mathbb{E}[\varepsilon]$.
\subsubsection{$\mathbb{E}[\varepsilon|Id]$}
First, we assume that packet $i-1$ finds the system in idle state and starts service. Therefore, we have $T_{i-1}=S_{i-1}$ and we assume the inter-arrival time between $i-1$ and next packet is $X$ (the next packet may be dropped and may not be indexed). If $X<S_{i-1}$, we have $Y_i=S_i$. Otherwise if $X>S_{i-1}$, we have $Y_i=X-S_{i-1}+S_{i}$. Now, we have:
\begin{align*}\vspace{-0.1in}
	\nonumber	\mathbb{E}[\varepsilon|Id]=\int_{0}^{H}\int_{s_{i-1}}^{\infty}\int_{H-x}^{\infty}&(x+s_{i}-H)\\
	\nonumber &f_{S}(s_i)f_{X}(x)f_{S}(s_{i-1})ds_idxds_{i-1}
\end{align*}\vspace{-0.2in}
\begin{align*}
\nonumber	+\int_{H}^{\infty}\int_{s_{i-1}}^{\infty}\int_{0}^{\infty}&(x+s_{i}-s_{i-1})\\
	\nonumber &f_{S}(s_i)f_{X}(x)f_{S}(s_{i-1})ds_idxds_{i-1}
\end{align*}\vspace{-0.2in}
\begin{align*}
	\nonumber+\int_{0}^{H}\int_{0}^{s_{i-1}}\int_{H-s_{i-1}}^{\infty}&(s_{i-1}+s_{i}-H)\\
	\nonumber &f_{S}(s_i)f_{X}(x)f_{S}(s_{i-1})ds_idxds_{i-1}
\end{align*}\vspace{-0.2in}
\begin{align}\label{EeMG12}
	+\int_{H}^{\infty}\int_{0}^{s_{i-1}}\int_{0}^{\infty}s_{i}
 &f_{S}(s_i)f_{X}(x)f_{S}(s_{i-1})ds_idxds_{i-1}.
\end{align}\vspace{-0.1in}
\subsubsection{$\mathbb{E}[\varepsilon|B]$}
Next, we assume that packet $i-1$ finds the system in busy state. We use $W$ to denote the waiting time for this packet in the buffer and we have the pdf for $W$ as: $f_W(w)=\frac{{\rm Pr}\{S>w\}}{\mathbb{E}[S]}$. Note that only when $X>W$, this packet will be indexed as the $i-1$th packet and $T_{i-1}=W+S_{i-1}$. If $W<X<W+S_{i-1}$, we have $Y_i=S_i$. Otherwise if $X>W+S_{i-1}$, we have $Y_i=X-W-S_{i-1}+S_{i}$. Now, we have:
\begin{align*}
	\nonumber	\mathbb{E}[\varepsilon|B]=&\\
	\nonumber \int_{0}^{H}&\int_{0}^{H-s_{i-1}}\int_{H-s_{i-1}-w}^{\infty}\int_{w}^{w+s_{i-1}}(w+s_{i-1}+s_{i}-H)\\
	\nonumber &\ \ \ f_{X}(x)f_{S}(s_i)f_{W}(w)f_{S}(s_{i-1})dxds_idwds_{i-1}
\end{align*}\vspace{-0.2in}
\begin{align*}	
	\nonumber +\int_{0}^{\infty}&\int_{H-s_{i-1}}^{\infty}\int_{0}^{\infty}\int_{w}^{w+s_{i-1}}s_{i}\\
	\nonumber &\ \ \ f_{X}(x)f_{S}(s_i)f_{W}(w)f_{S}(s_{i-1})dxds_idwds_{i-1}
\end{align*}\vspace{-0.2in}
\begin{align*}
\nonumber	+\int_{0}^{H}&\int_{0}^{H-s_{i-1}}\int_{w+s_{i-1}}^{\infty}\int_{H-x}^{\infty}(x+s_{i}-H)\\
	\nonumber &\ \ \ f_{S}(s_i)f_{X}(x)f_{W}(w)f_{S}(s_{i-1})ds_idxdwds_{i-1}
\end{align*}\vspace{-0.2in}
\begin{align}\label{EeMG12B}
	\nonumber +\int_{0}^{\infty}&\int_{H-s_{i-1}}^{\infty}\int_{w+s_{i-1}}^{\infty}\int_{0}^{\infty}(x-w-s_{i-1}+s_{i})\\
&\ \ \ f_{S}(s_i)f_{X}(x)f_{W}(w)f_{S}(s_{i-1})ds_idxdwds_{i-1}.
\end{align}
Finally, we have overage probability as  $P_o=\lambda(\mathbb{E}[\varepsilon|Id]p_{I}+\mathbb{E}[\varepsilon|B]p_{B})$ from equation (\ref{Po}).
\subsection{Average Overage}
Similarly, we evaluate $\mathbb{E}[Q|Id]$ and $\mathbb{E}[Q|B]$ based on (\ref{Qi}):
\begin{align*}
	\nonumber	\mathbb{E}[Q|Id]=\int_{0}^{H}\int_{s_{i-1}}^{\infty}\int_{H-x}^{\infty}&\frac{1}{2}(x+s_{i}-H)^2\\
	\nonumber &f_{S}(s_i)f_{X}(x)f_{S}(s_{i-1})ds_idxds_{i-1}
\end{align*}\vspace{-0.2in}
\begin{align*}
	\nonumber	+\int_{H}^{\infty}\int_{s_{i-1}}^{\infty}\int_{0}^{\infty}\biggl(\frac{1}{2}(x+&s_{i}-s_{i-1})^2\\
	\nonumber+(x+s_{i}-s_{i-1})(s_{i}-H)\biggr) &f_{S}(s_i)f_{X}(x)f_{S}(s_{i-1})ds_idxds_{i-1}
\end{align*}\vspace{-0.2in}
\begin{align*}
	\nonumber+\int_{0}^{H}\int_{0}^{s_{i-1}}\int_{H-s_{i-1}}^{\infty}&\frac{1}{2}(s_{i-1}+s_{i}-H)^2\\
	\nonumber &f_{S}(s_i)f_{X}(x)f_{S}(s_{i-1})ds_idxds_{i-1}
\end{align*}\vspace{-0.2in}
\begin{align}\label{EQMG12}
	\nonumber	+\int_{H}^{\infty}\int_{0}^{s_{i-1}}\int_{0}^{\infty}&\biggl(\frac{1}{2}s_{i}^2+s_{i}(s_{i-1}-H)\biggr)\\
	 &f_{S}(s_i)f_{X}(x)f_{S}(s_{i-1})ds_idxds_{i-1}.
\end{align}\vspace{-0.2in}
\begin{align*}
	\nonumber	\mathbb{E}[Q|B]=&\\
	\nonumber \int_{0}^{H}\int_{0}^{H-s_{i-1}}&\int_{H-s_{i-1}-w}^{\infty}\int_{w}^{w+s_{i-1}}\frac{1}{2}(w+s_{i-1}+s_{i}-H)^2\\
	\nonumber &\ \ \ f_{X}(x)f_{S}(s_i)f_{W}(w)f_{S}(s_{i-1})dxds_idwds_{i-1}
\end{align*}\vspace{-0.2in}
\begin{align*}
	\nonumber +\int_{0}^{\infty}\int_{H-s_{i-1}}^{\infty}&\int_{0}^{\infty}\int_{w}^{w+s_{i-1}}\biggl(\frac{1}{2}s_{i}^2+s_{i}(w+s_{i-1}-H)\biggr)\\
	\nonumber &\ \ \ f_{X}(x)f_{S}(s_i)f_{W}(w)f_{S}(s_{i-1})dxds_idwds_{i-1}
\end{align*}\vspace{-0.2in}
\begin{align*}
	\nonumber	+\int_{0}^{H}\int_{0}^{H-s_{i-1}}&\int_{w+s_{i-1}}^{\infty}\int_{H-x}^{\infty}\frac{1}{2}(x+s_{i}-H)^2\\
	\nonumber &\ \ \ f_{S}(s_i)f_{X}(x)f_{W}(w)f_{S}(s_{i-1})ds_idxdwds_{i-1}
\end{align*}\vspace{-0.2in}
\begin{align}\label{EQMG12B}
	\nonumber +\int_{0}^{\infty}\int_{H-s_{i-1}}^{\infty}&\int_{w+s_{i-1}}^{\infty}\int_{0}^{\infty}\biggl(\frac{1}{2}(x-w-s_{i-1}+s_{i})^2\\
	\nonumber &+(x-w-s_{i-1}+s_{i})(w+s_{i-1}-H)\biggr)\\ &f_{S}(s_i)f_{X}(x)f_{W}(w)f_{S}(s_{i-1})ds_idxdwds_{i-1}.
\end{align}
Then we have $\mathbb{E}[\Delta^o]=\lambda(\mathbb{E}[Q|Id]p_{I}+\mathbb{E}[Q|B]p_{B})$ from equation (\ref{Do}).
\subsection{Stale Update Probability}
Now, we evaluate ${\rm Pr^*}\{T_i>H\}$ as ${\rm Pr^*}\{T_i>H|Id\}p_{I}+{\rm Pr^*}\{T_i>H|B\}p_{B}$. Since the packet that arrives in idle state will be a delivered packet and its system time $T_{i}=S_{i}$, we have:
\begin{align}\label{EPMG12}
	{\rm Pr^*}\{T_i>H|Id\}=\int_{H}^{\infty}f_{S}(s_i)ds_i=1-F_S(H).
\end{align}
For the packets that arrive in busy state, only the packets with $X>W$ can be delivered packets, and their corresponding system time is $T_{i}=W+S_{i}$. So we have:
\begin{align}\label{EPMG12B}
\nonumber	{\rm Pr^*}\{T_i>H|B\}=\int_{0}^{\infty}\int_{H-s_{i}}^{\infty}\int_{w}^{\infty}f_{X}(x)f_{W}(w)&f_{S}(s_i)\\&dxdwds_i.
\end{align}
Therefore, we have $P_s=1-(p_{I}+p_{B_{1}})+{\rm Pr^*}\{T_i>H\}$ from equation (\ref{Ps}). 		
	\section{M/M/1 Queue}
The analysis for overage probability has been done in \cite{hu2021status} as violation probability. Therefore, we present only the analysis for average overage and stale update probability in the M/M/1 system. Note that in this system, all the packets are delivered packets since the buffer size is unlimited. The packets are served following the first come first serve discipline. The service time for each packet follows an iid exponential distribution with mean $\frac{1}{\mu}$. %Due to complexity of analysis, we don't explore the M/GI/1 queue here but leave it for future study.
\subsection{Average Overage}
Based on equation (6) in \cite{hu2021status}, we have the joint probability density function of $T_{i-1}$ and $Y_{i}$ as:
\begin{align*}
\nonumber	f_{T,Y}(y,t)&=\\&(\mu^2-\lambda\mu)e^{\lambda t-\mu y-\mu t}-\mu^2e^{-\mu y-\mu t}+\lambda \mu e^{-\lambda y-\mu t}.
\end{align*}
Therefore we have:
\begin{align}\label{EQMM1}
\nonumber	\mathbb{E}[Q]=\int_{0}^{H}\int_{H-t}^{\infty}\frac{1}{2}(y+t-H)^2f_{T,Y}(y,t)dydt\\+\int_{H}^{\infty}\int_{0}^{\infty}\biggl(\frac{1}{2}y^2+y(t-H)\biggr)f_{T,Y}(y,t)dydt,
\end{align}
and we have $\mathbb{E}[\Delta^o]=\lambda\mathbb{E}[Q]$ from equation (\ref{Do}). Closed-form expressions are shown in Appendix.\ref{Amm1}.
\subsection{Stale Update Probability}
Based on equation (4) in \cite{hu2021status}, we have the pdf of system time $T_{i}$ as:
\begin{align*}
	f_{T}(t)=\mu(1-\frac{\lambda}{\mu})e^{-\mu(1-\frac{\lambda}{\mu})t}.
\end{align*}
Since all the packets are delivered packets in this system, we have $p_d=1$ and ${\rm Pr^*}\{T_i>H\}$ as:
\begin{align}\label{PsMM1}
	{\rm Pr^*}\{T_i>H\}=\int_{H}^{\infty}f_{T}(t)dt=e^{-\mu(1-\frac{\lambda}{\mu})H}.
\end{align}
Therefore, we have $P_s={\rm Pr^*}\{T_i>H\}=e^{-\mu(1-\frac{\lambda}{\mu})H}$ from equation (\ref{Ps}).
	\section{Numerical Results}
In this section, we present numerical results for the three metrics. We also performed packet-based queue simulations offline for $10^6$ packets in order to verify the analytical results. We consider exponential distributed service times with $f_{S}(s)=\mu e^{-\mu s}$ and gamma distributed service times with $f_{S}(s)=\frac{\beta^\alpha}{\Gamma(\alpha)} \beta^{\alpha-1} e^{-\beta s}$ where $\alpha$ is the shape parameter. %{\color{blue} We show the closed form expression for M/M/1/1 and M/M/1 schemes in our technical report  \cite{techreport}.}
\begin{figure}[!t]
	\centering{
		\includegraphics[totalheight=0.265\textheight]{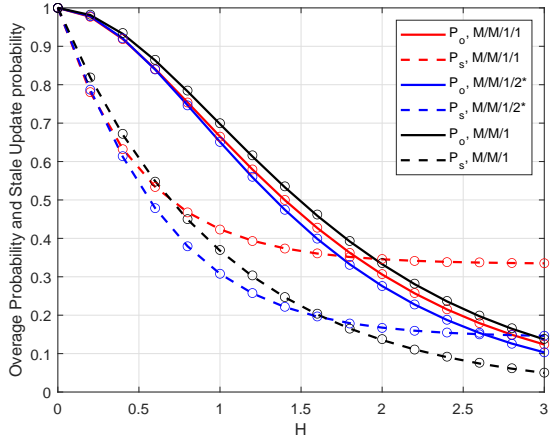}}\vspace{-0.15in}
	\caption{\sl Overage Probability and Stale Update Probability with respect to $H$ for $\lambda=1$ and exponential service time with $\mu=2$. Circles are simulation results.}
	\label{Num_1} 
\end{figure}

%{\color{red} Are all the symbols in the figures the simulation results? This needs to be mentioned then.}{\color{blue} What I want to mention here is that the numerical results have been verified with simulation result. All the figures here are generated with equation we show before.}

Fig. \ref{Num_1} shows the overage probability and stale update probability for the three queuing systems with respect to $H$ for $\lambda=1$ and exponential service time with $\mu=2$. We observe that when the threshold $H$ is small, $P_s$ is smaller than $P_o$ for all the three systems, but as $H$ increases, $P_s$ for M/M/1/1 and M/M/1/$2^*$ plateau, while the other probabilities keep decreasing. This is because the threshold has no effect on the queue behaviour, and with a large enough threshold, stale updates are mainly due to dropped packets in the system. Note that since the M/M/1 system serves all the packets, $P_s $ for M/M/1 keeps decreasing as the threshold increases.  
\begin{figure}[!t]
	\centering
	\subfigure[]{
		\includegraphics[totalheight=0.265\textheight]{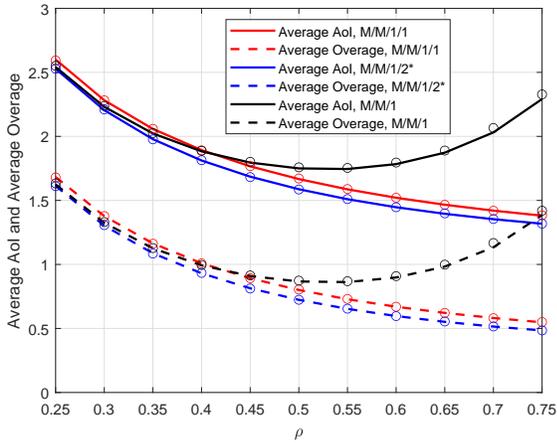}}
		\subfigure[]{
		\includegraphics[totalheight=0.265\textheight]{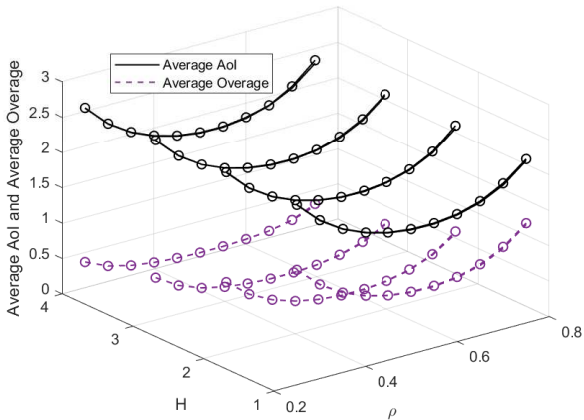}}\vspace{-0.15in}
	\caption{\sl Average AoI and Average Overage with respect to $\rho$ for exponential service time with fixed $\mu=2$. (a) All three queuing systems with $H=1$ (b) M/M/1 queue with respect to $\rho$ and $H$. Circles are simulation results.}
	\label{Num_2} 
\end{figure}

In Fig. \ref{Num_2}, we show average AoI and average overage with respect to $\rho$ where $\rho=\frac{\lambda}{\mu}$ for exponential service time with fixed $\mu=2$. In Fig. \ref{Num_2}(a), we compare the three queuing schemes with fixed $H=1$. It can be seen that the average overage has a similar behaviour to the average AoI with respect to $\rho$. For the M/M/1 queue, we can find optimum points for both average AoI and average overage while the average AoI and average overage keep decreasing as $\rho$ increases for M/M/1/1 and M/M/1/$2^*$. In Fig. \ref{Num_2}(b), we show average AoI and average overage for the M/M/1 queue with respect to $\rho$ and $H$. Note that the average overage is not actually equal to the average AoI minus $H$, because update packets with system times smaller than the threshold have zero overage. As $H$ becomes large, the average overage tends to zero, while the average AoI is unaffected by $H$, as can be seen in the figure. Further, the values of $\rho$ that minimize average AoI and average overage are slightly different.

%{\color{red} I added a sentence above.}

\begin{figure}[!t]
	\centering{
		\includegraphics[totalheight=0.265\textheight]{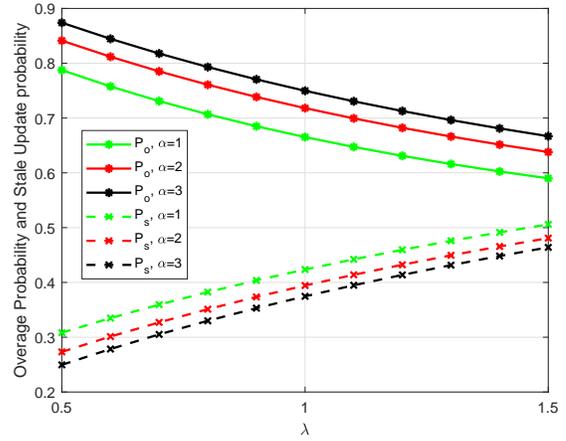}}\vspace{-0.15in}
	\caption{\sl Overage Probability and Stale Update Probability for M/GI/1/1 with respect to $\lambda$ for $H=1$ and gamma distribution service time with $\mathbb{E}[S]=\frac{1}{2}$.}
	\label{Num_3} 
\end{figure}
In Fig. \ref{Num_3}, we show Overage Probability and Stale Update Probability for M/GI/1/1 queue with respect to $\lambda$ for $H=1$ and gamma distributed service time with $\mathbb{E}[S]=\frac{1}{2}$. We can observe that with increasing $\lambda$, $P_o$ decreases while $P_s$ increases. This indicates that a trade-off exists for this system between overage probability and stale update probability. This is owing to the fact that as $\lambda$ increases, more packets are dropped in the queue, while the peak age decreases since the idle period decreases. We can also observe that as $\alpha$ increases, $P_o$ increases while $P_s$ decreases.

\begin{figure}[!t]
	\centering{
		\includegraphics[totalheight=0.265\textheight]{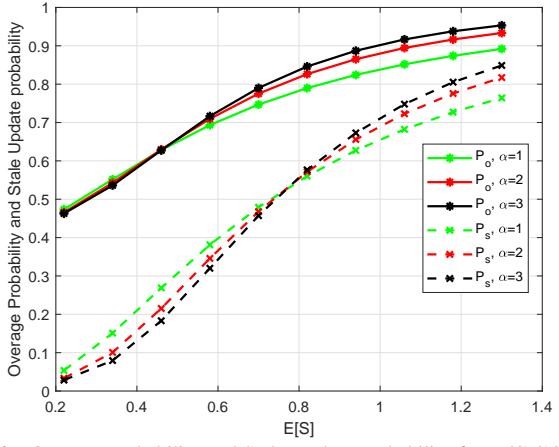}}\vspace{-0.15in}
	\caption{\sl Overage Probability and Stale Update Probability for M/GI/1/$2^*$ with respect to $\mathbb{E}[S]$ for gamma distribution service time and $H=1$, $\lambda=1$.}
	\label{Num_4} 
\end{figure}
Finally, in Fig. \ref{Num_4}, we show Overage Probability and Stale Update Probability for M/GI/1/$2^*$ queue with respect to $\mathbb{E}[S]$ for gamma distribution service time and $H=1$, $\lambda=1$. We can observe that as $\mathbb{E}[S]$ increases, both $P_o$ and $P_s$ increase. With smaller $\mathbb{E}[S]$, both $P_{o}$ and $P_{s}$ have good performance with larger $\alpha$ (i.e., smaller variance for service time); however, as $\mathbb{E}[S]$ increases, $P_{o}$ and $P_{s}$ are lower with smaller $\alpha$ (i.e., larger variance for service time). 
%{\color{red} Large $\alpha$ means large variance or small variance? The above sentence is not clear.}
This is because when the expected value of service time is small, a large variance implies that a large service time has a reasonably high probability of occurring, whereas when the expected value of service time is large, large variance means that small service times have a reasonably high probability of occurring.
	\section{Conclusion}
For a status update system with a soft deadline, this paper introduced three threshold-based age metrics -- {\em overage probability}, {\em average overage}, and {\em stale update probability} -- and evaluated them for three typical queuing systems. Expressions for the metrics are derived for three queueing systems, and several numerical results are obtained
%The {\em overage probability} is the probability that the age of the packet currently held by the receiver is larger than the threshold. The {\em stale update probability} is the probability that an update is stale, i.e., its age has exceeded the deadline, when it is delivered to the receiver. Finally, the {\em average overage} is defined as the time average of the update age beyond the threshold, and is a measure of the average ``staleness'' of the update packet held by the receiver. We intend to explore several new metrics and study their behavior in various systems. We investigate these three metrics in M/G/1/1, M/G/1/$2^*$ and M/M/1 queuing disciplines and we provide closed form expressions for these three metrics to compare them with AoI. 
%
Our numerical results show the behavior of these metrics under different parameter settings and different service time distributions. We compare the average overage with general average AoI to show their differences under different values of $\rho$. The results indicate a trade-off between these two for an M/M/1 queue. In the future, we intend to investigate these metrics for other queuing systems, and consider scheduling disciplines that minimize them.
	\appendix
\subsection{Closed form expressions for M/M/1/1}\label{Amm11}
In M/M/1/1 system, we have $f_S(s)=\mu e^{-\mu s}$ and $p_{I}=\frac{\mu}{\lambda+\mu}$.
\subsubsection{Overage Probability} \label{mm11op}
Based on (\ref{EeMG11}), we have:
\begin{align*}
	\nonumber	\mathbb{E}[\varepsilon|Id]=(\frac{1}{\lambda} + \frac{1}{\mu})e^{-\mu H} + \frac{He^{-\mu H}}{\lambda - \mu} + \frac{\mu^2(e^{-\lambda H} - e^{-\mu H})}{\lambda(\lambda - \mu)^2}
\end{align*}
and finally we have:
$P_o=\frac{\lambda\mu}{\lambda+\mu}\mathbb{E}[\varepsilon|Id]$.
\subsubsection{Average Overage}
Based on (\ref{EQMG11}), we have:
\begin{align*}
	\nonumber	&\mathbb{E}[Q|Id]=e^{-\mu H}(\frac{1}{\lambda^2}- H(\frac{1}{\lambda} + \frac{1}{\mu}) + \frac{1}{\mu^2} + \frac{1}{\lambda\mu}) \\
	&+ (He^{-\mu H} +\frac{e^{-\mu H}}{\mu})(\frac{1}{\lambda} + \frac{1}{\mu}) 
\\&+ \frac{\mu^2}{(\lambda - \mu)(\lambda - \mu)^3}\biggl((e^{-\lambda H}+ e^{-\mu H})	(\lambda^2 - 2\lambda\mu+ \mu^2 )H^2\end{align*}
\begin{align*}
&+(e^{-\lambda H}- e^{-\mu H})( 2\lambda - 2\mu)H\biggr)\\& -\frac{\mu^2H^2(e^{-\lambda H} - e^{-\mu H})}{(\lambda - \mu)^2} + \frac{\mu^2(e^{-\lambda H} - e^{-\mu H})}{\lambda^2(\lambda - \mu)^2} \\&- 2\mu^2 H\frac{e^{-\lambda H}-e^{-\mu H}(\mu H - \lambda H + 1)}{(\lambda - \mu)^3} +\frac{\lambda H e^{-\mu H}}{\mu(\lambda - \mu)}
\end{align*}
 and finally we have:
 $\mathbb{E}[\Delta^o]=\frac{\lambda\mu}{\lambda+\mu}\mathbb{E}[Q|Id]$
 \subsubsection{Stale Update Probability}
 Since we have:
 \begin{align*}
 	\nonumber	F_S(H)=1-e^{-\mu H}
 \end{align*}
 we can get $P_s=1-p_{I}F_S(H)=1-\frac{\mu}{\lambda+\mu}+\frac{\mu}{\lambda+\mu}e^{-\mu H}$.
\subsection{Closed form expressions for M/M/1/2$^{*}$}\label{Amm12}
In M/M/1/2$^{*}$ system, we have $p_{I}=\frac{\mu^2}{\lambda^2+\lambda\mu+\mu^2}, p_{B}=\frac{\lambda^2+\lambda\mu}{\lambda^2+\lambda\mu+\mu^2}$ and $p_{B_1}=\frac{\lambda\mu}{\lambda^2+\lambda\mu+\mu^2}$
\subsubsection{Overage Probability} 
Based on (\ref{EeMG12}) and (\ref{EeMG12B}), we have:
\begin{align*}
\mathbb{E}[\varepsilon|Id]&=He^{-H\mu }+\frac{e^{-H\mu }}{\mu }+\frac{e^{-H\lambda }-e^{-H\mu }}{\lambda }\\
&-\frac{e^{-H\mu }\left(e^{-H\lambda }-1\right)}{\lambda -\mu }+\frac{\mu e^{-H(\lambda+\mu)}}{\lambda ^2+\mu \lambda }\\
&+\frac{e^{-H\lambda }\left(1-e^{-H\mu }\right)}{\mu }+\frac{\lambda e^{-H\lambda }\left(e^{-H\mu }-1\right)}{\mu \left(\lambda -\mu \right)}
\end{align*}
\begin{align*}
\mathbb{E}[\varepsilon|B]&=\frac{e^{-H\mu} (1-e^{-H\lambda })}{\lambda }+\frac{ e^{-H(\lambda+\mu)}}{\lambda +\mu}\\
&-\frac{e^{-H\lambda }\left(e^{-H\mu }-1\right)}{\lambda }-\frac{e^{-H\lambda }\left(e^{-H\mu }-1\right)}{\mu }-He^{-H(\lambda+\mu)}\\
&+\frac{H\mu e^{-H\mu }}{\lambda }+\frac{\mu ^2e^{-H(\lambda+\mu)}}{\lambda {\left(\lambda +\mu \right)}^2}+\frac{2\mu e^{-H\mu }\left(e^{-H\lambda }-1\right)}{\lambda ^2}\\
&+\frac{H e^{-H(\lambda+\mu)}(\lambda-\mu)}{\lambda -\mu }+\frac{H\mu ^2e^{-H(\lambda+\mu)}}{\lambda ^2+\mu \lambda }\\
&+\frac{\lambda e^{-H\lambda }\left(e^{-H\mu }-1\right)}{\mu \left(\lambda -\mu \right)}-\frac{\mu e^{-H\mu }\left(e^{-H\lambda }-1\right)}{\lambda \left(\lambda -\mu \right)}
\end{align*}
Finally, we have overage probability as  $P_o=\lambda(\mathbb{E}[\varepsilon|Id]p_{I}+\mathbb{E}[\varepsilon|B]p_{B})$.
\subsubsection{Average Overage}
Based on (\ref{EQMG12}) and (\ref{EQMG12B}), we have:
\begin{align*}
\mathbb{E}[Q|Id]&=\frac{e^{-H\mu }}{\mu ^2}+H\frac{e^{-H\lambda }-e^{-H(\lambda+\mu)}}{\mu }+\frac{e^{-H\mu }\left(H\mu +1\right)}{\mu ^2}\\
&-\frac{e^{-H\lambda }\left(e^{-H\mu }-1\right)}{\mu ^2}+\frac{e^{-H(\lambda+\mu)}}{\lambda \left(\lambda +\mu \right)}+\frac{ e^{-H\lambda }-e^{-H(\lambda+\mu)}}{\lambda ^2}\\
&+\frac{\mu e^{-H(\lambda+\mu)}}{\lambda ^2\left(\lambda +\mu \right)}+\frac{ e^{-H\mu }-e^{-H(\lambda+\mu)}}{\mu \left(\lambda -\mu \right)}\\
&-\frac{\lambda \left(e^{-H\lambda }-e^{-H(\lambda+\mu)}\right)}{\mu^2 \left(\lambda -\mu \right)}+\frac{He^{-H\lambda }\left(e^{-H\mu }-1\right)}{\mu }\\
&+\frac{\mu e^{-H(\lambda+\mu)}\left(H\left(\lambda +\mu \right)+1\right)}{\lambda {\left(\lambda +\mu \right)}^2}-\frac{H\mu e^{-H(\lambda+\mu)}}{\lambda \left(\lambda +\mu \right)}
\end{align*}
\begin{align*}
&\mathbb{E}[Q|B]=\frac{2e^{-H\left(\lambda +\mu \right)}}{\mu \left(\lambda +\mu \right)}-\frac{2e^{-H\left(\lambda +\mu \right)}-2e^{-H\mu }}{\lambda\mu }-\frac{e^{-H\left(\lambda +\mu \right)}}{{\left(\lambda +\mu \right)}^2}\\
&+\frac{e^{-H\left(\lambda +\mu \right)}\left(H\left(\lambda +\mu \right)+1\right)}{{\left(\lambda +\mu \right)}^2}-\frac{2He^{-H\left(\lambda +\mu \right)}}{\lambda +\mu }-\frac{\mu e^{-H\left(\lambda +\mu \right)}}{{\left(\lambda +\mu \right)}^3}\\
&-\frac{\mu e^{-H\left(\lambda +\mu \right)}\left(H\left(\lambda +\mu \right)+1\right)}{{\left(\lambda +\mu \right)}^3}\\
&+\frac{e^{-H\left(\lambda +\mu \right)}\left(H\lambda ^3+3H\lambda ^2\mu +\lambda ^2+3H\lambda \mu ^2+3\lambda \mu +H\mu ^3+\mu ^2\right)}{\lambda ^2{\left(\lambda +\mu \right)}^2}\\
&+\frac{2e^{-H\mu }\left(e^{-H\lambda }-1\right)}{\lambda ^2}+\frac{He^{-H\mu }}{\lambda }+\frac{He^{-H(\lambda+\mu)}}{\lambda }\\
&-\frac{e^{-H\left(\lambda +\mu \right)}\left(\lambda -\mu -\lambda e^{H\lambda }+\mu e^{H\mu }+H\lambda ^2-H\mu ^2\right)}{\lambda ^2\left(\lambda -\mu \right)}
\end{align*}
and finally we have:
$\mathbb{E}[\Delta^o]=\lambda(\mathbb{E}[Q|Id]p_{I}+\mathbb{E}[Q|B]p_{B})$.
\subsubsection{Stale Update Probability}
Based on (\ref{EPMG12B}), we have:
\begin{align*}
{\rm Pr^*}\{T_i>H|B\}=\frac{\mu e^{-H(\lambda+\mu)}}{\lambda +\mu }-\frac{\mu e^{-H\mu }\left(e^{-H\lambda }-1\right)}{\lambda }
\end{align*}
we can get $P_s=1-(p_{I}+p_{B_{1}})+{\rm Pr^*}\{T_i>H\}$ where ${\rm Pr^*}\{T_i>H\}={\rm Pr^*}\{T_i>H|Id\}p_{I}+{\rm Pr^*}\{T_i>H|B\}p_{B}$.
 \subsection{Closed form expressions for M/M/1}\label{Amm1}
In M/M/1 system, the $\mathbb{E}[\varepsilon]$ is derived in \cite{hu2021status} as equation (29) and we have $P_o=\lambda\mathbb{E}[\varepsilon]$.  Next we show the closed-form expression for average overage in M/M/1.

\subsubsection{Average Overage}
Based on (\ref{EQMM1}), we have:
\begin{align*}
	\nonumber	\mathbb{E}[Q]&=	\frac{{e}^{-\mu H}}{\lambda ^2}-\frac{{e}^{-\mu H}}{\mu ^2}-\frac{{e}^{\lambda H-\mu H}}{\mu (\lambda -\mu)}\\&-\frac{{e}^{-\mu H }(\mu H +1)}{\mu ^2}-\frac{{e}^{-\mu H}({e}^{\lambda H}-1)}{\mu ^2}+\frac{H{e}^{\lambda H -\mu H }}{\lambda -\mu }\\&-\frac{{e}^{\lambda H -\mu H}(H\left(\lambda -\mu \right)-1)}{{(\lambda -\mu )}^2}-\frac{H{e}^{-\mu H }}{\lambda }\\&-\frac{\mu ({e}^{-\lambda H}-{e}^{-\mu H})}{\lambda ^2(\lambda -\mu)}+\frac{{e}^{-\mu H}({e}^{H\lambda }-1)}{\lambda \mu }+\frac{\lambda {e}^{\lambda H-\mu H}}{\mu ^2(\lambda -\mu)}\\&+\frac{{e}^{-\mu H }(\mu H+1)}{\lambda \mu }+\frac{\lambda {e}^{\lambda H -\mu H}(H(\lambda -\mu)-1)}{\mu {(\lambda -\mu)}^2}\\&-\frac{\lambda H {e}^{\lambda H -\mu H }}{\mu (\lambda -\mu)}	
\end{align*}
 and finally we have:
 $\mathbb{E}[\Delta^o]=\lambda\mathbb{E}[Q]$

	%\bibliographystyle{unsrt}  
	%\bibliography{references}

\end{document}